\documentclass[acmsmall]{acmart}% JEA uses acmsmall
\usepackage[binary-units,per-mode=symbol]{siunitx}
\renewcommand\footnotetextcopyrightpermission[1]{} % removes footnote with conference information in first column
\usepackage[utf8]{inputenc}
\usepackage{url}
\expandafter\def\expandafter\UrlBreaks\expandafter{\UrlBreaks%  save the current one
  \do\a\do\b\do\c\do\d\do\e\do\f\do\g\do\h\do\i\do\j%
  \do\k\do\l\do\m\do\n\do\o\do\p\do\q\do\r\do\s\do\t%
  \do\u\do\v\do\w\do\x\do\y\do\z\do\A\do\B\do\C\do\D%
  \do\E\do\F\do\G\do\H\do\I\do\J\do\K\do\L\do\M\do\N%
  \do\O\do\P\do\Q\do\R\do\S\do\T\do\U\do\V\do\W\do\X%
  \do\Y\do\Z}

\usepackage{booktabs}
\usepackage{pgfplots}
\usepackage{verbatim}
\usepackage{subfig}
\usepackage{xcolor}
\usepackage{soul}

\usepackage{amsthm}
\usepackage{algorithm}
\usepackage{multirow}
\usepackage{algorithmic}

\usepackage{graphicx}
\usepackage{epsf}
\usepackage{tikz}
\usepackage{tikzscale}
\usepgfplotslibrary{external}
\usetikzlibrary{external}
\usepackage{amsmath}

\usepackage{paralist}

\DeclareMathOperator{\xor}{\mathrm{~xor~}}
\pgfplotsset{compat=1.14}

\usetikzlibrary{positioning}

\setcopyright{acmlicensed}
\acmJournal{JEA}
\acmYear{2019} \acmVolume{1} \acmNumber{1} \acmArticle{1} \acmMonth{1} \acmPrice{15.00}\acmDOI{10.1145/3376122}
\begin{document}

\title{Binary Fuse Filters: Fast and Smaller Than Xor Filters}

%
% The "author" command and its associated commands are used to define the authors and their affiliations.
% Of note is the shared affiliation of the first two authors, and the "authornote" and "authornotemark" commands
% used to denote shared contribution to the research.
\author{Thomas Mueller Graf}
\email{thomas.tom.mueller@gmail.com}
%\orcid{1234-5678-9012}
\author{Daniel Lemire}
\orcid{1234-5678-9012}
\authornotemark[1]
\email{lemire@gmail.com}
\affiliation{%
  \institution{University of Quebec (TELUQ)}
  \streetaddress{5800 Saint-Denis, Office 1105}
  \city{Montreal}
  \state{Quebec}
  \country{Canada}
  \postcode{H2S 3L5}
}
% The footnote says "Authors' address", but it should be "addresses"

\begin{abstract}
Bloom and cuckoo filters provide fast approximate set membership while using little memory. 
Engineers use them to avoid expensive disk and network accesses. 
The recently introduced xor filters can be faster and smaller than Bloom and cuckoo filters. The xor filters are within 23\% of the theoretical lower bound in storage as opposed to 44\% for Bloom filters.
Inspired by Dietzfelbinger and Walzer, we build probabilistic filters---called \emph{binary fuse filters}---that are within 13\% of the storage lower bound---without sacrificing query speed.
As an additional benefit, the construction of the new binary fuse filters can be more than twice as fast as the construction of xor filters.
By slightly sacrificing query speed, we further reduce storage to within 8\% of the lower bound. 
We compare the performance against a wide range of competitive alternatives such as  Bloom filters, blocked Bloom filters, vector quotient filters, cuckoo filters, and the recent ribbon filters. Our experiments suggest that binary fuse filters are superior to xor filters.
\end{abstract}



\maketitle 
\thispagestyle{empty}
\section{Introduction}
Querying a database for the presence of a key may be expensive. To accelerate processing, we sometimes use \emph{probabilistic filters}.
A probabilistic filter indicates quickly whether a given key is present in a set while using less memory than the set itself. In exchange for its small size and high speed, the probabilistic filter allows a small probability that it may falsely report that a key is present while it is not (a \emph{false positive}). The false positives may be later pruned after checking against the actual set. If there are few false positives, the net result might be higher speed. To guarantee a bound of $\epsilon$ on the false-positive probability, we require at least $ \log_2 1/ \epsilon$~bits per key.

The conventional probabilistic filter is the Bloom filter~\cite{Bloom:1970:STH:362686.362692,lovett2010lower}. Bloom filters use at least 44\% more memory than the theoretical lower bound: $ 1.44 \log_2 1/ \epsilon$~bits per key. Furthermore, the lower the \emph{false-positive} probability, the more random access queries and hash-function computations are needed ($\approx \log_2 1/ \epsilon$): e.g., seven accesses are required for a 1\% false-positive probability.
Cuckoo filters~\cite{Fan:2014:CFP:2674005.2674994} are often superior to Bloom filters. They may require as few as two distinct cache-line accesses and they may use less memory than Bloom filters. E.g., they may use only 30\% to 40\% more memory than the optimal bound in the 12-bit and 16-bit cases.

Based on a proposal by Dietzfelbinger and Pagh~\cite{10.1007/978-3-540-70575-8_32}, we recently proposed the xor filter as a practical competitive alternative~\cite{10.1145/3376122}. It requires three random accesses, irrespective of the desired false-positive probability, and it is within 23\% of the optimal memory usage. Experimentally, we found that xor filters can be both smaller and faster than Bloom and cuckoo filters for many realistic scenarios. E.g., they can be twice as fast at query time as Bloom filters with a similar false-positive probability. By sacrificing query-time performance, we could further compress part of the xor filter, getting within 15\% of the theoretical lower bound: we called the resulting filters xor+. The space-saving feature of xor+ filters comes at a significant performance cost. 

We wondered: could we get much closer to the theoretical lower bound in storage while keeping the high performance of xor filters?
Building on theoretical work by Dietzfelbinger and Walzer~\cite{dietzfelbinger_et_al:LIPIcs:2019:11159}, we propose a novel practical approach, the \emph{binary fuse filters}. They are conceptually similar to xor filters, and they rely on nearly the same simple code. Yet for sets spanning at least ten thousand keys, 
they are preferable to xor filters. Indeed, they offer
 a significantly reduced size of $\approx $13\% over the theoretical bound (down from 23\% with xor filters),
while maintaining the same good query time.

A downside of xor and xor+ filters~\cite{10.1145/3376122} is their relatively slow construction time. They typically require more construction time (e.g., $2\times$) than competing alternatives. The binary fuse filters, in addition to their smaller size, also benefit from faster construction times. In some instances, the construction speed can be twice that of xor filters. % along with a much reduced temporary memory usage. 
In effect, binary fuse filters render xor filters practically obsolete.

Both the original xor filters and the new binary fuse filters are based on a 3-wise construction. To further reduce the size of the filter, we introduce 4-wise binary fuse filters. We achieve an overhead over the theoretical bound of only about 8\% in exchange for a modest increase in query time ($\approx$30\%) compared to xor filters. The net result is advantageous compared to xor+ filters for users who need to minimize space usage.

% Maybe surprisingly, the implementation of the binary fuse filters is nearly identical to that of xor filters. They require relatively little code, and they are easily ported to different programming languages.
\section{Related Work}

We are interested in fast and practical probabilistic filters with nearly optimal memory usage. A probabilistic filter is built from a set. Given a candidate key, it never fails to correctly report that the key belongs to the set, but it may sometimes erroneously report set membership, with a small probability. 
In practice, given a bound on the false-positive probability, probabilistic filters have a determined capacity at the time of the construction with a corresponding memory usage. 
Given a desired bound $\epsilon$ on the false-positive probability, we need at least $\log_2 1/\epsilon$~bits per key in the set in general.

Data structure  like hash tables or B-trees have a capacity 
and memory usage that can be increased or reduced over time.
Practical probabilistic filters offering bounded false-positive probability and nearly optimal memory usage usually have a fixed capacity determined at construction time. Some of these probabilistic filters, such as the xor filter, assume that all keys are available at construction time whereas others, like Bloom filters, allow a progressive construction. 
Some filters, such as counting Bloom filters and cuckoo filters,
allow removing keys.
 Constructing a new filter can be fast: many filters require less than \SI{100}{\nano\second} per key at construction time (see \S~\ref{sec:experiments}).
 We may construct a Bloom filter faster than we may sort the input data~\cite{10.1145/3376122}.
%\hl{Thomas: what about: "Other filters, like the counting Bloom filter, and the cuckoo filter, allow for removing keys later on."? We would need to add references of course.}
% \danielinline{Unless a referee demands that we address removal, I am not sure what this adds to the paper since this is not an issue that was address at all. And removal (with access to the original set) is only one of many other functionality that probabilistic filters have. You do have extensible Bloom filters, for example, and so forth.}
% Thomas: I agree!

The conventional probabilistic filter is the Bloom filter~\cite{Bloom:1970:STH:362686.362692}.
It is made of an array of bits and a collection of $k$~hash functions $h_1, h_2, \ldots, h_k$ that map  keys to bit indexes. When adding a key to the filter, we compute the $k$~hash functions and set the corresponding $k$~bits to 1. When doing a membership test, we check whether the values of the $k$~bits are set to 1. %Given a fixed number of hash functions and a predetermined array of bits, the false-positive probability increases as we add more keys to the filter.
Bloom filters use as little as
$1.44 \log_2 1/\epsilon$~bits per key, when setting $k = \log_2 1/\epsilon$. That is, Bloom filters use $\approx 44$\% more memory than the lower bound of $\log_2 1/\epsilon$~bits per key in the best case, when they are at full capacity.

Thus, Bloom filters require many independent random accesses in a bit array and much computation when the false-positive probability $\epsilon$ is small: e.g., we may need $k=14$~hash-function computations and random accesses per query for  $\epsilon = 1/\num{10000}$. 
Blocked Bloom filters~\cite{Putze:2010:CHS:1498698.1594230,lang2019performance} provide higher speed by requiring that all bits corresponding to a given key fit in a small block of bits. It comes at a significant storage penalty (e.g., 30\% more than Bloom filters) but can provide more speed by reducing cache misses and computation. State-of-the-art blocked Bloom filters also benefit from the fast SIMD (single-instruction-multiple-data) instructions available on commodity processors~\cite{Polychroniou:2014:VBF:2619228.2619234}. If reduced space is not a critical feature, block Bloom filters are competitive.

There is a broad family of alternatives to the Bloom filter approach relying instead on the concept of a \emph{fingerprint}. We choose a hash function that maps each 
potential key 
to a $k$-bit word. Given a set, we construct a data structure that maps each key of the set to its fingerprint. At query time, we compare the computed fingerprint with the retrieved fingerprint from the data structure. If they match, then we conclude that the key might be in the set. As long as the data structure acts as a map for the keys of the original set, there can be no false negative: any key in the set is correctly identified as being potentially in the set since the fingerprints are identical. If a key is not in the set, then the data structure could return any other fingerprint. We randomly pick the fingerprint-generating function from a universal family~\cite{carter1979universal} so that the probability that two distinct keys map to the same fingerprint is no more than $1/2^k$---our false-positive probability is $\epsilon = 1/2^k$. If we could construct a data structure that contained just $k$~bits per key in the original set, we would have an ideal probabilistic filter in the sense that it would have the smallest possible storage requirement given a false-positive bound. It is possible to get close to such a lower bound, but engineers often wish for fast construction and fast query speeds.

Several practical probabilistic filters based on fingerprints have been proposed.
\emph{Golomb-compressed sequences}~\cite{Putze:2010:CHS:1498698.1594230} store the sorted partial hash codes and fingerprints, encoded as differences between values. Golomb-compressed sequences have relatively poor query performance. \emph{Cuckoo filters}~\cite{Fan:2014:CFP:2674005.2674994} store fingerprints in a cuckoo hash table. At full capacity ($\approx 94\%$), they use less space than Bloom filters for a comparable false-positive probability, and they are typically faster at query time for low false-positive probabilities. There are other related competitive approaches such as quotient filters~\cite{Pandey:2017:GCF:3035918.3035963,PandeyCDB21} and 
Morton filters~\cite{Breslow:2018:MFF:3213880.3232248}.

In practice, a Bloom filter may be constructed with a given capacity for a given objective false-positive probability. When it contains fewer values than the set capacity, the false-positive probability will be lower. As more keys are added, the false-positive probability increases. 
In contrast, fingerprint techniques have generally a fixed false-positive probability. %Thus, we should avoid fingerprint-based filters with excess capacity if we seek to make the best use out of our storage.

In recent work, we proposed the xor filter as a new competitive fingerprint technique~\cite{10.1145/3376122}. It can be viewed as a reduced form of the \emph{Bloomier filter}~\cite{Chazelle:2004:BFE:982792.982797, charles2008bloomier}. The xor filter is made of an array of $k$-bit words. At query time, we map the provided key to three locations in the array, and we compute the bitwise xor of the three corresponding words. We check that it matches the fingerprint of the key. 
By following the algorithm from Botelho et al.~\cite{Botelho:2007:SSM:2394893.2394911}, we demonstrated practically that it was possible to map the fingerprints from a set of size $n$ to an array containing $\approx 1.23 \times n $~words---as long as $n$ is large enough. In effect, we leave about 23\% of the words to zero. Furthermore, we can further compress the array to try to recover some of the wasted (zero) space with an approach that we called the xor+ filter. However, the xor+ filter comes at a significant performance penalty since bit values cannot be directly accessed with a single memory lookup.
%\hl{TODO PRESENT RIBBON FILTERS HERE.}
%\hl{Thomas: what about (see also https://engineering.fb.com/2021/07/09/data-infrastructure/ribbon-filter/)
The even more recently proposed ribbon filters are also based on fingerprints,
but---unlike xor filters---use Gaussian elimination.
%They offer configurable space overheads and false-positive probabilities: we refer the interested reader to 
Dillinger and Walzer~\cite{Dillinger2021RibbonFP} find that though xor filters have better query-time performance, ribbon filters
have several advantages. Ribbon filters are more configurable than xor filters, 
supporting many different false-positive probabilities and they can trade
construction time for space efficiency. Ribbon filters also have
fast construction time compared to xor filters: our own experiments
support this observation.

\begin{verbatim}

\end{verbatim}

\section{Binary Fuse Filter}
\label{sec:binaryfusefilter}

The binary fuse filter is a variation on the xor filter~\cite{10.1145/3376122}. 
It is made of a fingerprint function, an array of fingerprints and a hash function from keys to locations in the array of fingerprints.
We review the construction of the xor filter. 

We start from a set of $n$~distinct keys from a given universe of keys. Furthermore, we need a fingerprint function that maps each possible value from the universe to a word value (e.g., an 8-bit word). The size of the word in bits determines the false-positive probability: each additional bit reduce the false-positive probability by half. The fingerprint function may be provided by the user. The construction itself consists of picking a hash function and populating an array of fingerprints.

We begin by describing a 3-wise construction though we also extend it to a 4-wise construction.
%We allocate an array containing slightly more than $n$~words. Conceptually, we want to build quickly an efficient perfect hash function~\cite{cichelli1980minimal} which assigns each distinct set key to exactly one key in the array. We want an injective function: each key in the array either does not correspond to any set key (it is \emph{wasted}), or else it corresponds to exactly one set key. We want to build this map in a manner that is amenable to an efficient probabilistic filter. 
We pick three hash functions mapping each key of the set to three distinct locations in the array. These hash functions are picked at random from a family of hash functions. For each key, we want to pick one \emph{primary} location out of the three array locations. Furthermore, we seek an ordering of the keys such that 
 no key is mapped to the primary location of an earlier key.
% \hl{Thomas: instead of "an earlier key", what about writing "any other key"? I think it would be simpler. Sure, it would allow for a different technique (we use peeling) but in this summary maybe it's OK?}

The xor filter is built out of such a construction. To construct an xor filter, we pick an array, which contains $\approx 1.23 n$~words, and we divide it into three nearly equal segments (of size $\approx 1.23 n/3$).\footnote{The ratio 1.23 between the size of the array and the size of the set holds asymptotically for $n$ large.} Given a 3-way hash function from the set to the array, each word in the array is mapped from $\approx 3/1.23 \approx 2.4$~keys of the set, on average. 
We can count how many keys from the set are mapped to each location.
We can scan the array, and when we find a location that corresponds to exactly one key from the set, we match this key to this location. We remove the matched key from consideration as if it never existed: we decrement the three counters corresponding to its corresponding three distinct locations. Each time we remove a key, we potentially create new locations corresponding to a single key. With high probability, this \emph{peeling} process eventually terminates when all keys have been removed. 
As a result, we have constructed an injective function. In case of failure, we restart the process with different hash functions---again picked at random. 
The exact success probability depends on the size of the set and on the size of
the array: we found empirically that it was greater than 90\% when
the array contains $1.23 n + 32$~words~\cite{10.1145/3376122}. For large sets, the probability converges to 100\%. 
We can bound the failure probability and thus the expected construction time: a lower bound of 90\% on the success rate implies that the expected number of trials is no more than $\approx 1.1$.
We can then build an xor filter as follows. We start from the last matched key, and we iterate backward through all  keys---in the reversed order that they were matched.
Each time, we query the three word values in the array corresponding to the key, and modify the value in the array at the one location corresponding to the key so that the \emph{bitwise exclusive or} (xor) value of the three word values is equal to the fingerprint of the key. By construction, once we have read an array value, we never modify it again. Thus, if we pick any key from the set, and xor the three corresponding values from the array, we find again the fingerprint of the key.

To query an xor filter, starting from a potential key, we compute the three hash functions. We access the three corresponding locations in the array. We compute the \emph{bitwise exclusive or} value of the three values, and we compare it with the fingerprint of the potential  key. If the key is part of the set, then the fingerprint matches the aggregate of the three values. If the key is not part of the set, then it is likely that its fingerprint  differs from the aggregate.

An xor filter is fast at query time. Although we query three distinct locations in the array to check any membership, modern processors can sustain ten, twenty or more memory loads in parallel at any one time per core. However, the construction can be relatively slow since we rely extensively on unpredictable random accesses. As the size of the array grows, our construction struggles to make good use of the fast cache of the processor. We can slightly optimize the construction with buffers, but we have been unable to make the construction of the xor filter as fast as that of a comparable Bloom filter.

The xor filter relies on a three-way partition of the array. Dietzfelbinger and Walzer~\cite{dietzfelbinger_et_al:LIPIcs:2019:11159} suggest an approach that might be preferable, especially for larger values of $n$: we can divide the array into many small same-size non-overlapping segments (e.g., hundreds) and then have the hash function pick three distinct locations corresponding to three \emph{consecutive} segments.\footnote{Walzer~\cite{walzer2021peeling} examined a variation where the segments are overlapping.}
They note that this construction works well for large values of $n$.
For small values of $n$, we can set the number of segments to three, in which case
the construction, and the space usage, match the xor filter.
Hashing keys to a power-of-two integer range is easier and faster than hashing to an arbitrary range.
Thus, to ease the computation, we further require that the segments span a power of two. Hence, the generation of the hash function may only involve the selection of a starting segment followed by the efficient computation of three hash values within a power-of-two range. In this manner, we hash keys to three locations within three consecutive segments. 
Except for this difference in the choice of the hash functions, everything is conceptually the same as for xor filters during the construction and the queries.
We intuitively improve locality: a set value is mapped to a smaller range of addresses in memory. It also allows us to use a smaller array while still having a low failure probability. For large sets, we only need an array with $\approx 1.125 n$~elements (see Table~\ref{table:formula}): a space reduction of nearly 10\% compared to the xor filter. Following Dietzfelbinger and Walzer's terminology, we name this approach a \emph{fuse filter}. We further qualify it as \emph{binary} because of the requirement that the disjoint segments span a power of two.

So far, we have limited the discussion to three hash functions. By replacing the three hash function with four hash function, we can reduce even more the size of the array, making it closer to $n$. The construction is the same. We map each key to four (instead of three) locations in the array. 
We find in practice that for large sets (e.g., containing one million keys), we only need an array of size $1.075 n$. It is a modest space saving of about 5\% compared to the three-wise fuse filter, but a saving of $\approx 15$\% compared to the xor filter. We call the resulting construction a \emph{4-wise fuse filter}.

%\hl{Thomas: we use "arity" here, but one referee prefers "3-wise" over "3-ary", so not sure if "arity" needs to be replaced as well? If yes I don't know how...}
In theory, we could increase further the arity to achieve further space savings, as suggested by Dietzfelbinger and Walzer~\cite{dietzfelbinger_et_al:LIPIcs:2019:11159}, but we are already within 7.5\% of the theoretical lower bound with the 4-wise construction. Furthermore, each increase in the arity comes with a computational cost. In the other direction, decreasing the arity increases the storage requirements. In effect, the cuckoo filter could be viewed as a practical realization of the 2-wise scenario, although it requires the replacement of simple array locations with buckets for efficiency.

\begin{figure}
%\subfloat[false-positive probability from $10^{-10}$\%]{
%\includegraphics[width=0.48\textwidth]{spaceusage.tikz}
%}
%\subfloat[false-positive probability from $2^{-16}$ to $2^{-8}$]{
\includegraphics[width=0.48\textwidth]{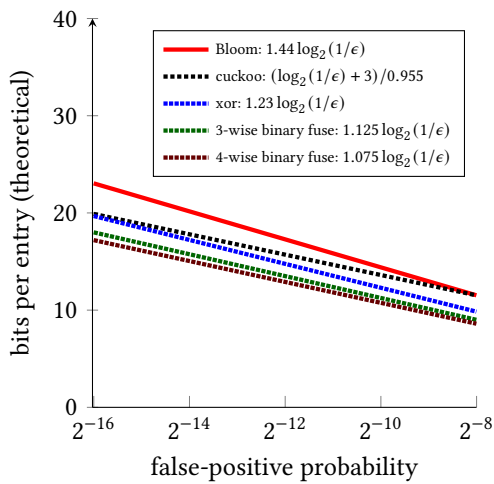}
%}
\caption{\label{fig:spaceusagetheory} Theoretical memory usage for Bloom filters (optimized for space), cuckoo filters (at maximal capacity~\cite[Table~2]{Fan:2014:CFP:2674005.2674994}), xor filters and binary fuse filters given a desired bound on the false-positive probability from $2^{-16}$ to $2^{-8}$. }
\end{figure}

In Fig.~\ref{fig:spaceusagetheory}, we compare theoretical bounds of various probabilistic filters including the binary fuse filters. We omit xor+ filters which require
$1.0824 \log_2(1/\epsilon) +0.5125$~bits per entry: strictly worse than 4-wise binary fuse filters ($1.075 \log_2(1/\epsilon)$).
Over the selected range ($2^{-16}$ to $2^{-8}$), the 4-wise fuse filter is the most economical.
For very small false-positive probabilities, a cuckoo filter becomes slightly smaller.
For example, the 3-wise and 4-wise fuse filter require 
65~bits and 62~bits respectively to achieve the same 
false-positive probability as a 64-bit cuckoo filter ($3\times 10^{-16}$\%).

%\hl{Thomas: I tried to combine (a) and (b) into one range}
%\begin{figure}
%\subfloat[false-positive probability from $2^{-16}$ to $2^{-6}$]{
%\includegraphics[width=0.48\textwidth]{spaceusage-combined.tikz}
%}
%\caption{\label{fig:spaceusagetheory-combined} Theoretical memory usage for Bloom filters (optimized for space), cuckoo filter (at maximal capacity~\cite[Table~2]{Fan:2014:CFP:2674005.2674994}), xor filters and binary fuse filters given a desired bound on the false-positive probability. }
%\end{figure}

\begin{figure}
\includegraphics[width=0.48\textwidth]{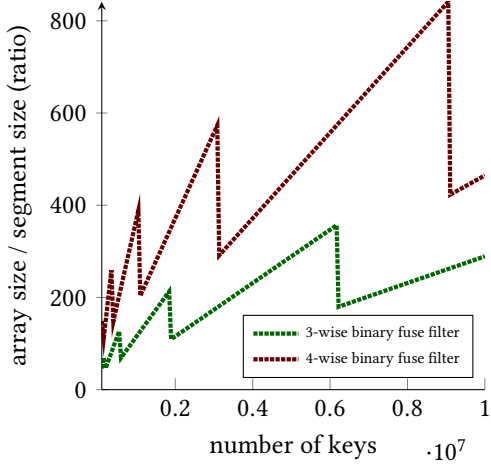}
\caption{\label{fig:ratiotheory} Ratio between the array size and the segment sizes for various number of keys in binary fuse filters. }
\end{figure}

We find it interesting that it is possible to build a binary fuse filter much faster than an xor filter. The main speed limitation when building an xor filter over a large set is related to the random accesses. It takes much time, for example, to load and update three counter values at three random locations in a large array. A binary fuse filter
can be constructed with better memory locality as follows. See Algorithm~\ref{alg:construction}.
\begin{itemize}
    \item A given key is mapped to consecutive segments in the array (e.g., three segments in the 3-wise case). In a first pass through the input keys, we partially sort them to a second buffer by the segment they are mapped to. This can be done efficiently, in linear time with a single pass, because there are few segments (e.g., 300).
    \item We can then scan through the partially sorted keys and update a temporary array of counters---such an array tells us how many keys map to a given location. Because the keys are sorted by segment, a forward pass through the keys tends to access the counters in a forward manner, thus reducing the number of cache misses compared to a random approach.
    \item We scan the array of counters to identify the locations corresponding to a single key. The corresponding keys are added to a stack. Since we scan forward, the later keys in the stack tend to correspond to later locations.
    \item We then unwind the stack, virtually removing the keys, and conditionally adding the newly isolated keys. These marked keys are added to another stack. This process tends to visit the locations backwards, from the last locations to the first locations. Usually, the process terminates successfully.
    \item  Finally, starting from the latter stack, we can construct the binary fuse filter, by making sure that the bitwise exclusive or of the values at the three locations map to the fingerprint of the keys. By construction, we tend to go from locations that are at the beginning of the array, working toward the end of the array.
\end{itemize}
In effect, given that the disjoint segments are hundreds of times smaller than the overall array (see Fig.~\ref{fig:ratiotheory}), we naturally tend to work in a local manner which makes it easier for the system to keep the needed values inside the CPU cache. The only additional effort is the initial partial sort, but it is relatively inexpensive.

Algorithm~\ref{alg:construction} relies on an array of sets. In practice, we implement the sets using 
an integer-value counter and a fixed-length mask initialized at zero. When adding a key, we increment the counter and
compute the exclusive-or of the key with the mask, storing the result as the new mask. We similarly remove a key by decrementing
the counter and computing an exclusive-or. When the set is a singleton, the mask contains the element value.

\begin{algorithm}
\caption{Filter Construction for 3-wise binary fuse filters\label{alg:construction}.}
\begin{algorithmic}
\REQUIRE a set containing  $n$~(distinct) keys $S$ from universe $U$
\STATE Pick a fingerprint function $F$ from $U$ to $k$-bit integers for some value of $k>0$ which  determines the false-positive probability ($\epsilon = 1/2^k$).
\STATE Allocate $H \leftarrow$ an array of size $\approx 1.125 n$ containing $k$-bit integers (initialized at zero),  divided into segments of size $2^{\lfloor  \log_{3.33} n + 2.25\rfloor } $ (see Table~\ref{table:formula}).
\STATE
\REPEAT
\STATE Pick hash functions $h_0, h_1, h_2$ from 
$U$ to array locations in $H$ so that $h_0(x), h_1(x), h_2(x)$ occupy three distinct and consecutive segments.
\STATE Sort the keys $x$ in set $S$ by the index of the segment containing $h_0(x)$: the first keys should map to the first segment, then the next segment and so forth.
\STATE Allocate an array $c$ of sets having the same size as $H$. The sets are initially empty.
\STATE Scan through the keys $x$ in set $S$ in order, and add $x$ to the sets $c[h_0(x)], c[h_1(x)], c[h_2(x)]$. 
\STATE  Scan through the array $c$ in order, and each time a singleton is found, append the location to a stack ($Q$).
\STATE  Initialize an empty stack $P$
\WHILE{stack $Q$ is not empty}
\STATE Pop a location $i$ from the stack $Q$
\IF{$c[i]$ is a singleton}
\STATE Append the key $x$ contained in $c[i]$ and its location $i$ to the stack $P$
\STATE Remove the key $x$ from lists $c[h_0(x)], c[h_1(x)], c[h_2(x)]$. If any of these sets become singleton, add their location to stack $Q$.
\ENDIF
\ENDWHILE
\UNTIL{the size of $P$ is $n$}

\STATE 

\WHILE{stack $P$ is not empty}
\STATE Pop a key $x$  and a location $i$ from the stack $P$.  We have that $i\in \{h_0(x),h_1(x),h_2(x)\}$.
\STATE Set $H[i] \leftarrow H[h_0(x)] \xor H[h_1(x)] \xor H[h_2(x)] \xor F(x)$.  
\ENDWHILE
\STATE  \textbf{return} $H$ as well as the corresponding hash functions $h_0, h_1, h_2$
\end{algorithmic}
\end{algorithm}

Table~\ref{table:formula} provides the formulas  to compute the size of the fingerprint array and the size of the segment sizes. For large sets, the array sizes tend toward $1.125 n$ (3-wise) and $1.075n$ (4-wise) according to these formulas.
In practice, we further round up the fingerprint array size so that it is a multiple of the  segment size. We also ensure that there are at least 3~segments in the 3-wise case, and at least 4~segments in the 4-wise case, but that is only a concern for tiny sets.
We arrive at these parameters empirically after building simulations. We find that the failure probability, when using these formulas for sets ranging from a few keys to hundreds of millions, are less than 1\%---typically far lower than 1\%.
\begin{table}
\caption{Initial fingerprint array sizes and segment sizes given a set of size $n$. The actual fingerprint array sizes are rounded up so that they are divisible by the segment sizes.} 

\label{table:formula}

% \danielinline{The reason the formulas are a bit longer is that so you can plot them in your head because $0.875 + 0.25=1.25$ and $0.77 + 0.305=1.075$. I don't mind \emph{simplifying} them, but I am not sure it is a net win as far as clarity goes. We should avoid giving the impression that we just made up a formula. }

\centering
\begin{tabular}{ccc}\toprule
     & array size & segment size \\\midrule
% 3-wise    & 
% $\lfloor \max \left (1.125, 0.875 + \frac{3.45}{ \ln n}\right ) n \rfloor $ 
% & $2^{\lfloor \frac{\ln n}{ \ln 3.33} + 2.25\rfloor }$
% \\
 3-wise    & 
 $\left\lfloor \left ( 0.875 + 0.25 \max \left (1, \frac{\log 10^6}{\log n} \right )\right ) n \right\rfloor  \geq \lfloor 1.125 n \rfloor$ 
 & $2^{\lfloor  \log_{3.33} n + 2.25\rfloor } \approx 4.8\cdot n^{0.58}$ \\[1.5em]
  4-wise    & $\left\lfloor \left (0.77 + 0.305\max \left(1,\frac{\log (6 \cdot 10^5)}{\log n}  \right)\right ) n\right\rfloor \geq   \lfloor 1.075 n \rfloor$  &  $2^{\lfloor \log_{2.91}n  - 0.5\rfloor } \approx 0.7\cdot n^{0.65}$\\\bottomrule
\end{tabular}
\end{table}

\section{Experiments}
\label{sec:experiments}

%\hl{Thomas: In Table 1, instead of 1000000 and 600000 which are hard to read, what about using ${10^6}$ and ${6 \cdot 10^5}$ instead? Or $\num{1 000 000}$ and $\num{600 000}$?}

We test our algorithms using freely available C++ software.\footnote{\url{https://github.com/FastFilter/fastfilter_cpp}} 
 We support both Linux and macOS platforms.  
We use the GCC~10 compiler with the compiler flags \texttt{-O3}. 
Under the x64 systems, we add the  \texttt{-march=native} flag for the benefits of the blocked Bloom filter---which requires AVX2 instructions.
All software is single-threaded. We never access the disk or the network during the benchmarks.

Our primary test platform is an x64 Linux server with an Intel 
processor with Skylake microarchitecture:
an Intel i7-6700 processor running at \SI{3.4}{\GHz} GHz, 
with \SI{8}{\mebi\byte} of L3 cache.
For each filter, we run 3 tests, and report the median. 
Our error margin is less than 3\%.

Our secondary test platform is an x64 Linux server with an  AMD EPYC 7262 (Rome Zen~2, 2019) processor running at \SI{3.39}{\GHz}. It has \SI{32}{\kibi\byte} of L1 data cache, \SI{512}{\kibi\byte} of L2 cache and \SI{128}{\mebi\byte} of L3 cache.
Here, our error margin is much higher: 25\%,
and this is why it is not our primary test platform.
We expect that run-to-run variations are mostly due to caching effects,
which seem to be much higher here than for the older Skylake processors.
We ruled out frequency throttling: our benchmarking tool reports the effective processor frequency under Linux and the frequency variations are small (less than 1\%). 
Because of the higher error margin, we run 15~tests for each filter, 
and report the median.
The results are reported in appendix~\ref{appendix:benchmarkResultsAMD}.

Though our benchmarking software is written in C++, 
we have ported the binary fuse filters to other programming languages such as Java, C and Go, and we make them freely available.\footnote{\url{https://github.com/FastFilter/}} We verified that our good results are not specific to our C++ implementations.

For performance reasons, both our xor filters~\cite{10.1145/3376122} and binary fuse filters rely whole-byte fingerprint sizes (e.g., 8-bit, 16-bit). We do not expect our filters to be practical in terms of speed for arbitrary word sizes. We exclude xor+ filters from consideration: they require
more storage than  4-wise binary fuse filters and they are slower than xor filters both at query and construction time (e.g., by a factor of two).
When constructing xor or binary fuse filters, we do not work directly on the input keys 
during construction, but rather on their hashed values: therefore, the construction 
performance is not sensitive to the choice of the keys in the provided set.

We compare our results against several competitive filters. Implementation details are important, and a missed optimization can provide a misleading picture. However, we make all our implementations freely available online.

\begin{itemize}
\item We implemented the standard Bloom filter algorithm with configurable  false-positive probability and size. We test with 12 and 16 bits per key, and the respective number of hash functions $k$ that are needed for the lowest false-positive probability. For fast construction and membership tests, we rely on double hashing:  we hash only once with a 64-bit function, treated as two 32-bit values $h_1(k)$ and $h_2(k)$. The  hash functions are $g_i(k) = h_1(k) +i \cdot h_2(k) \bmod 2^{32}$ for $i=0,\ldots,k-1$. We use an optimized  construction algorithm that hashes the keys in bulk, within batches, before setting the bits in the array---thus improving memory-level parallelism. Thus, our implementation of Bloom filters is not naive: we attempt to provide a sensible baseline.
\item We use a highly optimized blocked Bloom filter from Apache Impala. It was designed for the Intel AVX2 256-bit instruction set. Queries require access to a single cache line. We have extended the implementation so that it works under processors not supporting AVX2 instructions---but the results are not equally impressive. Hence, we only benchmark the blocked Bloom filter with AVX2 instructions enabled.
\item We modified the original cuckoo filter implementation~\cite{githubCuckoofilter}. 
The maximum load was reduced from 0.96 to 0.94: a recommended workaround suggested by the cuckoo-filter authors. We further modified the code to lift the restriction to a power-of-two capacity~\cite{Fan:2014:CFP:2674005.2674994} since it may otherwise leave  too  much excess capacity.
We use 12-bit and 16-bit fingerprints.
\item We use the original implementation of ribbon filters~\cite{Dillinger2021RibbonFP} which was made available online as a GitHub fork of our own benchmarking code.\footnote{\url{https://github.com/pdillinger/fastfilter_cpp}}  We modified the implementation of the ribbon filters so that they use the same hashing algorithm as all other tested algorithms. It ensures that the construction of the filters is not practically sensitive to input values. There is a wide range of ribbon filters, and an exhaustive survey is beyond our scope. We use both the standard and homogeneous ribbon filters (8-bit and 15-bit versions).
\item We use a slightly modified implementation of vector quotient 
filters~\cite{PandeyCDB21} which was made 
available online.\footnote{\url{https://github.com/splatlab/vqf}}
We optimized the implementation to speed up both 
construction and lookup by replacing remainder instructions with
multiply-and-shift routines~\cite{LEMIRE2021e07442}. 
The relative maximum load was set to 0.93, to ensure 
reliable successful construction. 
We slightly changed the calculation of the fingerprint (\emph{tag}) 
so that the false-positive probability is consistent with the expectation for 8-bit
fingerprints ($1/2^8$).
On our test platforms, we rely on AVX2 instructions: AVX-512 instructions are unavailable.
%\hl{What about we also write (here or maybe later) something like this:
%"This limitation reduces the speed.
%On the other hand, the vector quotient filters are the only
%algorithm in our tests that rely on the pdep instruction, 
%which is only fast in Intel processors."
%or 
%"For highest performance, vector quotient filters require
%a CPU supporting fast select, as well as AVX-512 instructions.
%Our test platforms does supports select (via pdep instruction), but only AVX2: AVX-512 instructions are unavailable."
%-- what I want to say is, one could argue it's not fair we don't have AVX-512, on the other other hand only VQF uses pdep which is not 'fair' for the other algorithms, as they don't need it.
%}
We considered speeding up construction by pre-sorting the entries, but found 
empirically that doing so would reduce the maximum load to about 0.85. 
\end{itemize}

Our benchmark begins with a set of random 64-bit integers as keys. 
In practice, engineers might have sets of variable-length strings or other values, but we assume that all such objects can be hashed to 64-bit integers. Such hashing might cause collisions---when two values hash to the same value---which would increase the false-positive probability. Yet 64-bit hashing should have a negligible effect on the false-positive probability ($\approx 1/2^{64}$ times the size of the set).\footnote{We assume that $n/2^{64} \ll \epsilon$ where $n$ is the size of the set.} 
In our implementation,
we make no assumption about the distribution of the keys other than they should be distinct: we check that our results are  similar when, for example, we use sequential keys (e.g., $1, 2, 3, \ldots$).
For all implementations, we use a randomly seeded 64-bit murmur hash function~\cite{ivanchykhin2017regular} to compute the fingerprint from the key.

A common use case for a probabilistic filter is a large number of candidate keys that are not, in fact, in the set. Thus, we may want to use a query set that is distinct from the filtered set. Yet there  may be other use cases. Thus, we generate large  random query sets with 100~million keys containing a variable fraction of keys from the filtered set. % (0\%, 25\%, 50\%, 75\%, 100\%)
We select the scenario where 25\% of the keys are in the set.
We shuffle the query sets so that keys that are in the set may appear at any location---this should prevent the branch predictor in the processor from being unrealistically accurate.

\begin{figure}[htb]
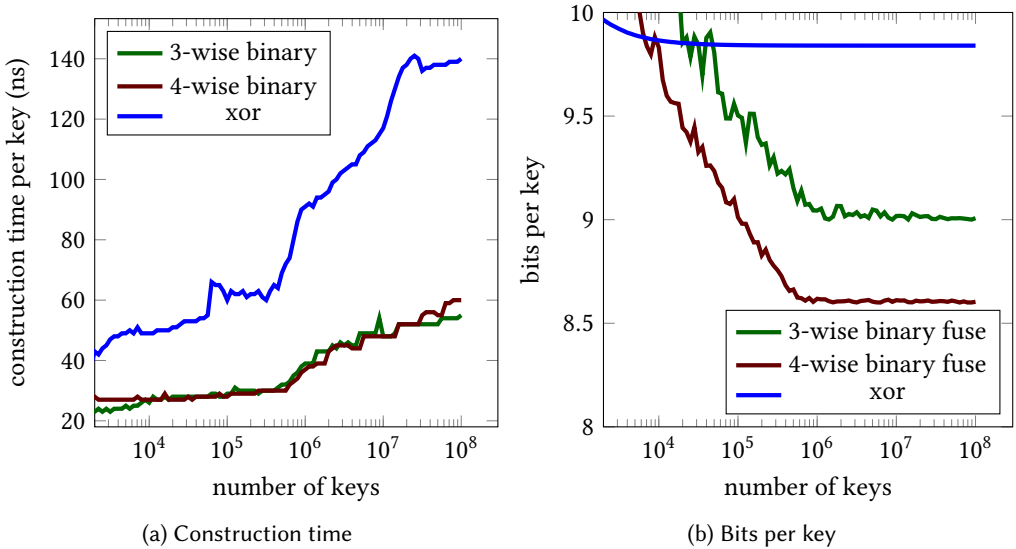

\subfloat[Construction time\label{fig:comparingxorbincons}]{
\includegraphics[width=0.48\textwidth]{constructime.tikz}
}
\subfloat[Bits per key\label{fig:comparingxorbinbits}]{
\includegraphics[width=0.48\textwidth]{bits.tikz}
}
\caption{\label{fig:comparingxorbin} 
Comparison between  8-bit xor and binary fuse filters for various
number of keys. The number of bits per entry for the 16-bit filters would be  double.}
\end{figure}
In Fig.~\ref{fig:comparingxorbin}, we present the construction time per key and the number of bits per key for variable numbers of keys in the filtered set. 
All these filters are based on an 8-bit construction and have an expected false-positive probability of $1/2^8\approx 0.4$\%. We compute the construction time  by repeating the construction as many times as necessary so that the total duration is at least \SI{0.1}{\second}. When more than one construction is needed, we compute the average time. We find that the construction time of both the 3-wise and 4-wise binary filters are similar and more than twice as fast as the construction of a comparable xor filter (Fig.~\ref{fig:comparingxorbincons}). The construction of the xor filter is not naive: it is an optimized construction and uses buffers to improve memory-level parallelism. The memory usage of the binary fuse filters is predicted by the formula in Table~\ref{table:formula}. Nevertheless, it is maybe useful to plot the actual memory usage of our implemented filters (Fig.~\ref{fig:comparingxorbinbits}). Recall that we round up the array sizes so that we have an integer number of segments so that we do not use exactly
%$\left \lfloor  0.875 + 0.25 \max \left (1, \log_n 1000000\right )  \right \rfloor $~bytes per key.
%\hl{Thomas: 1000000 is a bit hard to read... using spaces here with "num" doesn't make it easier I think; what about we use the following instead?}
$\left \lfloor  0.875 + 0.25 \max \left (1, \log_n 10^{6}\right )  \right \rfloor $~bytes per key.

As expected, experimentally, we find that for large sets---with at least a million  keys---the 3-wise filter reaches 9~bits/key, or 12.5\% worse than the theoretical minimum, while the 4-wise reaches about 8.6~bits per key, or 7.5\% worse than the theoretical minimum. These memory usages are preferable to the xor filter which uses 9.84~bits per key or 23\% worse than the theoretical minimum. For smaller sets, the benefits of the binary fuse filters are less significant. For tiny sets (fewer than \num{20000} keys), xor filters might even be slightly smaller due to a smaller overhead.\footnote{More expensive constructions~\cite{10.1007/978-3-319-38851} could be advantageous for small sets.}

Our benchmark consists of first constructing the probabilistic filter. And then, after picking a given query set, we repeatedly check whether keys of the query set are in the set according to the filter. We count the number of matches in a tight loop. To avoid undue compiler optimizations, we prevent the inlining of the \texttt{contain} function calls using the \texttt{noinline} function attribute under GCC\@. We find that our results are accurate with a run-to-run standard error of about 1\% for filtered sets containing 100~million keys and for query sets containing 10~million keys.

\begin{figure}[htb]\centering
\includegraphics[width=0.6\textwidth]{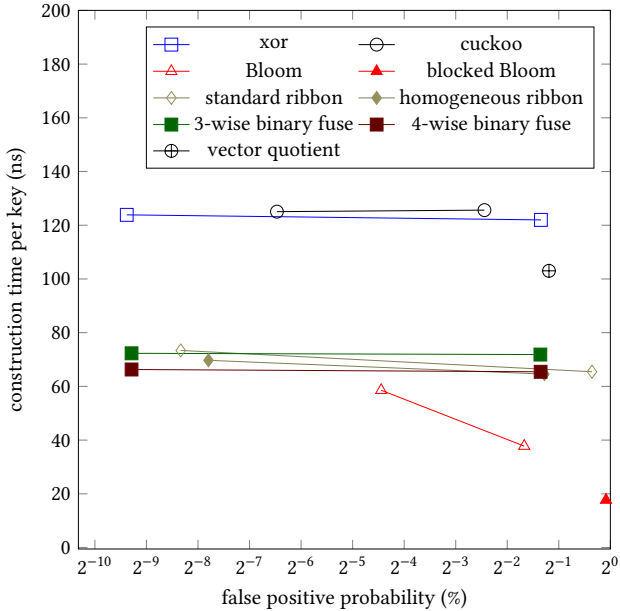}
\caption{\label{fig:constructiontime100m} Construction time versus the false-positive probability~(\%) for some families of filters on an Intel Skylake system with 100~million keys.  The lines between pairs of marks are for improved readability: only the marks represent actual data.}
\end{figure}

In Fig.~\ref{fig:constructiontime100m}, we plot the construction time per key. For filters that have a parameter determining the false-positive probability, we plot all values with lines between them: e.g., with xor filters, we include both the 8-bit and 16-bit versions. The fastest construction is with the blocked Bloom filter, but it has a relatively high fast-positive probability. The binary fuse
as well as the ribbon filters offer the fastest construction for filters that have less than 20\% space overhead and a low false-positive probability. We find that xor and cuckoo filters have slower construction. The vector quotient filters
have faster construction than cuckoo filters.

\begin{figure}[htb]
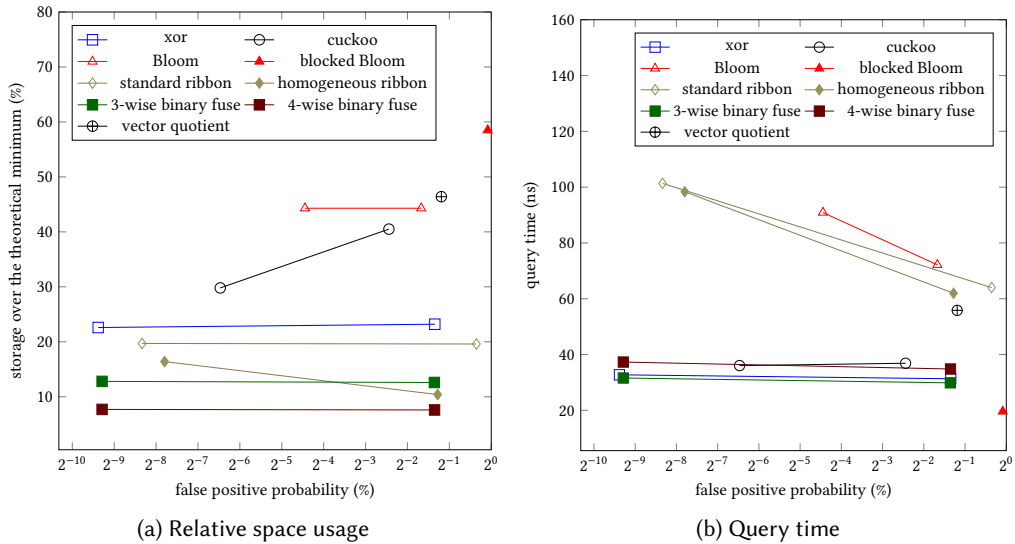

\subfloat[Relative space usage\label{fig:storage100m}]{
\includegraphics[width=0.48\textwidth]{spacevsfpp100M.tikz}
}
\subfloat[Query time\label{fig:querytime100m}]{
\includegraphics[width=0.48\textwidth]{query-timevsfpp100M.tikz}
}
\caption{
 Relative space usage compared to theoretical minimum and
query time (25\% in-set) versus the false-positive probability~(\%) for some families of filters on an Intel Skylake system with 100~million keys. The lines between pairs of marks are for improved readability: only the marks represent actual data.}
\end{figure}

In Fig.~\ref{fig:storage100m}, we plot the  memory usage of the various filters normalized against the theoretical minimum: $\log_2 1/\epsilon$ where $\epsilon$ is the measured (effective) false-positive probability. A relative measure makes comparisons between filters having different false-positive probabilities easier. The blocked Bloom filter has the worst memory efficiency, followed by the Bloom and vector quotient filters. 
The standard and homogeneous ribbon filters as well as  the binary fuse filters are more economical in our tests.   

% \danielinline{Here I qualified the statement regarding  ribbon filters (standard and homogeneous).}
%\hl{Thomas: what about we add "Note we only tested a small segment of the all possible filter variants" or something like that? Because we left out ribbon filters that have a very small storage overhead.}

In Fig.~\ref{fig:querytime100m}, we present the query-time performance. The fastest filter is the blocked Bloom filter. We find that the xor and 3-wise binary fuse filters have the next best performance. The 4-wise  binary fuse filter and cuckoo filters have similar albeit slightly worse query performance in our tests. The Bloom filters are slower, followed by the ribbon and vector quotient filters.

By comparing Fig.~\ref{fig:constructiontime100m} and Fig.~\ref{fig:querytime100m}, we find that most filters have a query time (per key) that is within a factor of two of the construction time (per key). In our experiments, only the xor and cuckoo filters are several times more expensive to construct than to query.

\section{Conclusion}

In many applications, we must index a static set at high speed, and using little space. Though Bloom filters remain a competitive solution, they use 44\% more space than the theoretical lower bound while practical solutions such as 3-wise binary fuse filters use far less (only $\approx 12$\%).  While our new work on binary fuse filters makes our recent work on xor filters in large part obsolete, we should draw the reader's attention to ribbon filters which perform well in our experiments.\footnote{For an exhaustive analysis and experimental assessment of ribbon filters, we refer the interested reader to Dillinger and Walzer~\cite{Dillinger2021RibbonFP}.} 
%%Furthermore, if space is not a major concern, blocked Bloom filters are an excellent choice in terms of construction and query speed.

Xor and binary fuse filters require 
access to the full set of keys at construction time. In this sense, they are immutable. 
Alternatives have typically a fixed memory usage and a maximal capacity, but they also
allow more flexibility such as progressive construction (adding keys one
by one). This observation suggests at least two directions for future
research. On the one hand, we could design more flexible algorithms to build
binary fuse filters: e.g., so that they become bulk updatable. 
On the other hand, we could seek to better document applications 
where immutable probabilistic filters are best suited: e.g., 
when the filter is sent over a network, bundled with a front-end or
used in a multi-threaded environment where locking is undesirable~\cite{5476244,Li:2014:AIF:2592798.2592820}.

Our experimental evaluation is single-threaded.  In networking or database applications, filters must be capable of high concurrency.
Future work should assess binary-fuse filters performance on multiple cores with a focus on  memory bandwidth, core-level memory parallelism
and realistic tasks. We also did not consider advanced SIMD-based optimizations such as those made possible by
the AVX-512 instruction sets~\cite{10.14778/3303753.3303757}.

In our software implementation, we rely on randomized murmur hash functions~\cite{ivanchykhin2017regular} to construct our filters.
However, theoretical analysis often relies on fully random hash functions~\cite{walzer2021peeling,dietzfelbinger_et_al:LIPIcs:2019:11159}.
Future research could seek to bridge this gap between theory and practice: fully random hash functions are typically impractical.

\begin{acks}
We are grateful to Apple for his initial code contributions. His work encouraged us to pursue this research. We are also thankful to Walzer for bringing to our attention their work~\cite{dietzfelbinger_et_al:LIPIcs:2019:11159}.
We thank Dillinger for pointing out a randomness flaw in our earlier benchmark framework.
\end{acks}

\bibliographystyle{ACM-Reference-Format}
\bibliography{binaryfusefilter}

\appendix

\section{Benchmark Results on AMD}
\label{appendix:benchmarkResultsAMD}

\begin{figure}\centering
\subfloat[Construction time\label{fig:constructiontime100m-rome}]{
\includegraphics[width=0.49\textwidth]{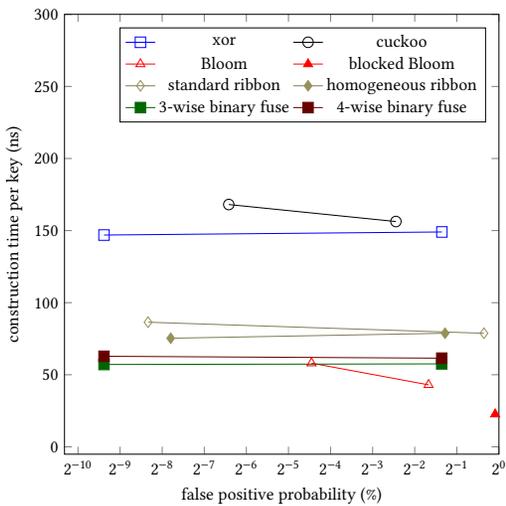}
}
\subfloat[Query time\label{fig:querytime100m-rome}]{
\includegraphics[width=0.49\textwidth]{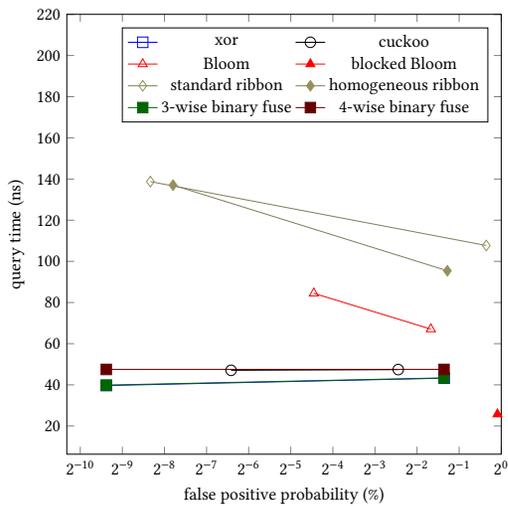}
}

\caption{Construction and query time versus the false-positive probability~(\%) for some families of filters on an AMD Rome system with 100~million keys.  The lines between pairs of marks are for improved readability: only the marks represent actual data.}
\end{figure}

In Fig.~\ref{fig:constructiontime100m-rome}, we plot the construction time per key. We find that on this platform, binary fuse filters are faster to construct than ribbon filters. In Fig.~\ref{fig:querytime100m-rome}, we plot the query time per key.
We find that ribbon filters are slightly worse, relatively speaking,
compared to our primary test platform. 
The vector quotient implementation relies on the \texttt{pdep} instruction which is slow
on this AMD processor: it makes  the vector quotient implementation non- competitive
and we omit it from consideration.

\end{document}